\documentclass[onecolumn,11pt,a4paper,showpacs,preprintnumbers,amsmath,amssymb]{revtex4}
\date{March 2010}
\usepackage{epsfig}
\usepackage{latexsym}
\usepackage{amssymb}
\usepackage{booktabs}
\usepackage{graphicx}
\newcommand{\be}{\begin{equation}}
\newcommand{\ee}{\end{equation}}
\newcommand{\ba}{\begin{eqnarray}}
\newcommand{\ea}{\end{eqnarray}}
\newcommand{\bi}{\begin{itemize}}
\newcommand{\ei}{\end{itemize}}
\newcommand{\ud}{\,\mathrm{d}}
\newcommand{\tr}{{\rm Tr\,}}
\newcommand{\re}{\mathop{\rm Re}}

\newcommand{\nn}{\nonumber \\}

\newcommand{\half}{{\textstyle\frac{1}{2}}}

\newcommand{\<}{\langle}
\renewcommand{\>}{\rangle}
\newcommand{\eq}{Eq.~}

\newcommand{\fig}{Fig.~}

\newcommand{\la}{\label}

\newcommand{\txts}{\textstyle}

\newcommand{\dBdloga}{\frac{d\beta}{d\log a}}

\newcommand{\im}{\mathop{\rm Im}}
\newcommand{\as}{a_{\sigma}}
\newcommand{\at}{a_{\tau}}
\newcommand{\betas}{\beta_{\sigma}}
\newcommand{\betat}{\beta_{\tau}}

\newcommand{\Ss}{S_{\sigma}}
\newcommand{\St}{S_{\tau}}

\newcommand{\Nt}{N_{\tau}}
\newcommand{\Ns}{N_{\sigma}}

\newcommand{\by}{\boldsymbol{y}}
\newcommand{\bx}{\boldsymbol{x}}
\newcommand{\bp}{\boldsymbol{p}}
\newcommand{\bq}{\boldsymbol{q}}
\newcommand{\bh}{\boldsymbol{h}}

\begin{document}
\title{The transverse structure of the QCD string}

\author{Harvey~B.~Meyer}
\affiliation{Johannes Gutenberg Universit\"at Mainz, 
    Institut f\"ur Kernphysik, 55099 Mainz, Germany}

\date{\today}

\begin{abstract}
The characterization of the transverse structure of the QCD string 
is discussed. 
We formulate a conjecture as to how the stress-energy tensor 
of the underlying gauge theory couples to the string degrees of freedom.
A consequence of the conjecture is that the energy density and the 
longitudinal-stress operators measure the distribution of the transverse 
position of the string, to leading order in the string fluctuations, 
whereas the transverse-stress operator does not.
We interpret recent numerical measurements of the transverse size of the 
confining string and show that the difference of the energy and 
longitudinal-stress operators is the appropriate probe to 
use when comparing with the next-to-leading order string prediction.

Secondly we derive the constraints imposed by open-closed string duality 
on the transverse structure of the string. 
We show that a total of three independent `gravitational' 
form factors characterize the transverse profile of the closed string,
and obtain the interpretation of recent effective string theory calculations:
the square radius of a closed string of length $\beta$ 
defined from the slope of its gravitational form factor,
is given by $\frac{d-1}{2\pi\sigma}\log\frac{\beta}{4r_0}$ in $d$ space dimensions.
This is to be compared with the well-known result that 
the width of the open-string at mid-point grows as 
$\frac{d-1}{2\pi\sigma}\log \frac{r}{r_0}$.
We also obtain predictions for transition form factors
among closed-string states.
\end{abstract}

\maketitle

\section{Introduction}
The area law of Wilson loops in lattice gauge theories~\cite{Wilson:1974sk}
has long been interpreted in terms of a string formation 
by the flux lines. 
In SU(3) gauge theory the area law, $\<W\>\sim e^{-\sigma A}$, 
signals the linear confinement of heavy quarks $Q$ and $\bar Q$: 
the static potential takes the form $V(r)\sim \sigma r$, 
where $\sigma$ is identified with the string tension.
Once the quark-interdistance $r$ is significantly larger than 
the confinement scale $\sqrt{\sigma}$, it was realized 
a long time ago that the corrections to the static potential, 
as well as the low-energy excitations of the $Q\bar Q $ system,
could be described by an effective two-dimensional theory~\cite{Luscher:1980fr}.
This `worldsheet' theory of the 
$d-1$~\footnote{$d$ is the number of spatial dimensions.}
massless degrees of freedom $\bh$,
namely the transverse fluctuations of the string, led to two
important predictions: firstly, the linear potential receives
a $1/r$ corrections, the L\"uscher term~\cite{Luscher:1980ac}, 
and its excitations are spaced by $\frac{\pi}{r}$ gaps. Secondly, the 
amplitude of the transverse string fluctuations grows logarithmically
with the length of the string~\cite{Luscher:1980iy},
\be
w^2_{lo}\equiv \<\bh^2\> =\frac{d-1}{2\pi\sigma}\log\frac{r}{r_0}.
\la{eq:wlo}
\ee
It is this second aspect of the low-energy string dynamics that is 
the focus of this paper.

Recently, highly accurate numerical results have been obtained 
in the $d=2$ SU(2) gauge theory
for the expectation value of local operators in the presence of 
a static $Q\bar Q$ pair~\cite{Gliozzi:2010zv}. 
Measured as a function of the distance $|\by|$ from the $Q\bar Q$ axis, 
is defines a distribution whose second moment was successfully 
compared to the effective-theory prediction (\ref{eq:wlo}).
In view of these results and of the prospect of pushing the 
comparison to next-to-leading order, the first issue we wish to address
is the precise connection between the profile 
probed by a local gauge-theory operator and the worldsheet
expectation value of $\bh^2$.

\section{Coupling of the stress-energy tensor to the confining string\la{sec:coupling}}
With the physical picture of a fluctuating `thin' string in mind,
the stress and energy stored in the flux lines 
is entirely carried by the string. 
If only the transverse component of a string element's motion 
contributes to the string energy,
then the Hamiltonian is $H= \int \frac{dm}{\sqrt{1-v_\perp^2}}$, 
where $dm$ is the rest mass of an element of the string.
For the Nambu-Goto string, $dm=\sigma ds$, where $ds$ is the 
length of the string element, but more sophisticated possibilities,
such as a curvature term, should be kept 
in mind~\cite{Polyakov:1986cs}\footnote{In $d=2$, a curvature
contribution would come in the form $dm=\sigma ds
[1+\frac{\partial_1^2 \bh}{(1+(\partial_1\bh)^2)^2}]$, 
where $ds= dy_1\sqrt{1+(\partial_1\bh)^2}$.}.
Retaining only the simplest rest mass contribution, the expression 
for the string energy density in Minkowski space reads,
in the static gauge,
\be
T_{00}(t,y_1,\by)={\sigma}\,
\frac{1+(\partial_1\bh)^2}
{\sqrt{1+(\partial_1\bh)^2-(\partial_t\bh)^2 
-(\partial_t\bh)^2(\partial_1\bh)^2
+(\partial_t\bh\cdot \partial_1\bh)^2}}
\,\delta^{d-1}(\by-\bh(t,y_1)).
\la{eq:T00}
\ee
Here $\bh$ is the worldsheet field; it has $d-1$ components.
The worldsheet is parametrized by $t$ and $y_1$, the coordinate that runs
along the $Q\bar Q$ axis, and $\by$ contains the $d-1$ coordinates 
transverse to the string. The worldsheet indices are denoted generically
by $a,b,\dots$
We thus expect the energy-density operator of the underlying 
gauge theory to couple to the worldsheet operator appearing in this 
expression, understood as an expansion in $\partial_a \bh$.
At zeroth order, we have simply 
$T_{00}(t,y_1,\by) = \sigma \delta^{d-1}(\by-\bh(t,y_1))$, 
which means that the distribution measured by the energy density
operator coincides with the distribution in $\bh$.
In particular the second moments in $\by$
of the transverse distribution obtained from
$T_{00}$ is expected to match the expression for $\<\bh^2\>$ 
calculated in the worldsheet theory.
Expanding \eq(\ref{eq:T00}), we obtain
\be
T_{00} = \sigma\delta(\by-\bh)\Big[1+\half((\partial_t\bh)^2+(\partial_1\bh)^2)
-{\txts\frac{1}{8}}[(\partial_1\bh)^2]^2 
+{\txts\frac{3}{8}}[(\partial_t\bh)^2]^2
+{\txts\frac{1}{4}}(\partial_1\bh)^2(\partial_t\bh)^2
-{\txts\frac{1}{2}} (\partial_t\bh\cdot \partial_1\bh)^2\Big].
\la{eq:T00exp}
\ee
This expression suggests that at leading order in the fluctuations,
$\int d^{d-1}\by \,\by^2\, T_{00}(t,y_1,\by)$ measures the worldsheet
expectation value of the operator
\be
 \bh^2(1+\half (\partial_t\bh)^2+\half(\partial_1\bh)^2).
\ee
Thus when comparing  Monte-Carlo data for $\int d^{d-1}\by \,\by^2\, T_{00}(t,y_1,\by)$
with the effective theory, the worldsheet 
expectation value of $\bh^2$ needs to computed to next-to-leading order,
a tour-de-force achieved very recently~\cite{Gliozzi:2010zt}, 
but also the leading order expectation 
value of $\bh^2 \half ((\partial_t\bh)^2+(\partial_1\bh)^2)$
needs to be calculated. It is probably simpler to work 
with the operator $T_{00}-T_{11}$, for which we will see that 
the undesirable contribution of the quadratic fluctuations cancels out
(\eq\ref{eq:T00T11}).

It is instructive to note that the energy density expression
(\ref{eq:T00}) derived from geometric considerations coincides 
with the form of the canonical energy density derived from the L\"uscher-Weisz 
worldsheet action with the standard Noether procedure. Indeed the NLO Lagrangian reads
\be
{\cal L}^{\rm ws} = \half \partial_c\bh\cdot \partial^c\bh
+ c_2 (\partial_a\bh\cdot \partial^a\bh)(\partial_b\bh\cdot \partial^b\bh)
+ c_3 (\partial_a\bh\cdot \partial_b\bh)(\partial^a\bh\cdot \partial^b\bh)+\dots
\ee
with \emph{a priori} free coefficients $c_2$ and $c_3$,
and the stress-energy tensor 
\be
T^{\rm ws}_{ab}=\partial_a\bh\cdot \partial_b\bh 
+ 4c_2 (\partial_c\bh\cdot \partial^c\bh)\,(\partial_a\bh \cdot \partial_b \bh)
+4c_3 (\partial_a\bh\cdot \partial_c\bh)\,(\partial_b\bh\cdot \partial^c\bh)
-g^{ab} {\cal L}^{\rm ws},
\la{eq:Tab}
\ee
with in particular 
\be
T^{\rm ws}_{00}(t,y_1)=
 \half((\partial_t\bh)^2+(\partial_1\bh)^2)
+(c_2+c_3) ((\partial_1\bh)^2-3(\partial_t\bh)^2)
+2c_2 (\partial_t\bh)^2(\partial_1\bh)^2
+2c_3 (\partial_t\bh\cdot \partial_1\bh)^2 .
\la{eq:T00ws}
\ee
Expressions (\ref{eq:T00ws}) and (\ref{eq:T00}) are consistent for 
$c_2=\frac{1}{8}$ and $c_3=-\frac{1}{4}$, which are the Nambu-Goto 
values~\footnote{In fact, the quartic terms are already proportional 
to those in (\ref{eq:T00}) for $c_3=-2c_2$, a relation already found 
to play a special role in~\cite{Luscher:2004ib}.}.

In $d=2$ space dimensions, it was shown in~\cite{Luscher:2004ib} 
that these values for the `low-energy constants' $c_2$ and $c_3$
are the only ones compatible with open-closed string duality.
It was subsequently shown that this requirement is also equivalent 
to requiring that closed string have a relativistic dispersion relation, 
in other words requiring Poincar\'e invariance~\cite{Meyer:2006qx}.
If one requires that the effective string theory also describes 
a situation where the worldsheet itself is a torus
in a way that is consistent with the open- and closed-string spectral 
representations, then these values are the only ones possible 
in any dimension~\cite{Aharony:2009gg}.
In view of the geometric interpretation of the energy density operator,
these results show that only a string that is `immaterial', 
i.e.~for which only transverse motion of an element of the string
contributes to the string energy, yields a spectrum that is consistent
with open-closed string duality. Were it not for this fact, 
the fraction in \eq(\ref{eq:T00}) would have been replaced 
by $\sqrt{\frac{1+\partial_1\bh^2}{1-(\partial_t\bh)^2}}$,
which in particular does not yield
mixed term $(\partial_t\bh\cdot \partial_1\bh)^2$.
In other words, numerical evidence that the open string spectrum
requires $c_2$ and $c_{3}$ to take up their respective Nambu-Goto 
values really confirms the `immaterial' nature of the confining string.

It is also of interest to write out the 
expressions for the longitudinal stress operator $T_{11}$ explicitly 
(see \eq(\ref{eq:Tab})),
\ba
T_{11}(t,y_1,\by) = \sigma \delta(\by-\bh(t,y_1))
&\!\!\Big[\!\!& -1 + \half((\partial_t\bh)^2+(\partial_1\bh)^2)
+(c_2+c_3) (3(\partial_1\bh)^2-(\partial_t\bh)^2)
\nn && 
-2c_2 (\partial_t\bh)^2(\partial_1\bh)^2
-2c_3 (\partial_t\bh\cdot \partial_1\bh)^2 \Big].
\la{eq:T11}
\ea
This formula implies that the transverse string profiles obtained with 
$T_{00}$ and $T_{11}$ differ at quadratic order in $\bh$.
There is a specific reason why $T_{11}$ is an interesting probe 
of the string profile.
The transverse profile of the open string depends in general
at what point $y_1$ along the string it is measured. 
It is easy to see that if one uses $T_{11}$,
then the total longitudinal stress inside a transverse spatial slice, 
$\int \ud^{d-1}\by T_{11}(y_1,\by)$, does not depend on the position
$y_1$ of the slice along the string. This is simply because from 
the closed-string point of view, $T_{11}$ plays the role of the 
energy density operator, and therefore its forward matrix elements
are diagonal in an energy-eigenstate basis.
Evaluated on the ground state, this integrated longitudinal stress
yields the static force, 
\be
\int \ud^{d-1}\by\, \<T_{11}(y_1,\by)\>_{Q\bar Q} = 
-\frac{\partial E_0(r)}{\partial r}
\qquad (\forall y_1).
\la{eq:T11sr}
\ee
Because of this distinguishing property, the integrated longitudinal stress
is conserved along the open string, and it is natural to ask how the 
transverse distribution of longitudinal stress changes 
as one moves along the string.

It is however not true than any operator tracks the 
movement of the string at leading order. 
Take for instance the transverse operator $T_{22}$.
There is no corresponding operator on the worldsheet, since it is 
a two-dimensional field theory. 
One can show that 
\ba
\int dy_1\int d^{d-1}\by\, \<T_{22}(y_1,\by)\>_{Q\bar Q} &=& 0,
\ea
when correlated with the pair of Polyakov loops. 
The physical reason why the three-point function of $T_{22}$ 
vanishes is that the string does not, on average, exert any stress 
along the transverse directions.
Because the sum rule of this operator does not yield a term
proportional to the length of the string, this operator is not 
measuring, to leading order in the fluctuations $\bh$, 
the position of the string.
Therefore one cannot define a transverse distribution of the string
with a probabilistic interpretation based on this operator.
Instead this operator is sensitive in leading order to the expectation value of 
higher-derivative worldsheet operators.

We have followed the approach of L\"uscher and Weisz~\cite{Luscher:2004ib}
and worked in the static gauge. The point of view adopted by Polchinski and 
Strominger~\cite{Polchinski:1991ax} puts more emphasis on the conformal 
symmetry of the worldsheet theory, which severely constrains the class of 
actions they consider. It is therefore worthwhile to investigate the 
fate of conformal symmetry in the static gauge as well. This issue is left 
for a future study. We simply note that the trace of 
the canonical energy-momentum tensor
\be
T^{{\rm ws},\,a}_{\,a} = 2c_2 (\partial_a \bh \cdot \partial^a \bh)^2
+2c_3 (\partial_a\bh\cdot \partial_b\bh)(\partial^a\bh\cdot \partial^b\bh)
+\dots
= 2{\cal L}^{(4)}+\dots
\ee
no longer vanishes at the quartic order. However it is well-known 
that the canonical energy-momentum tensor is in general not traceless
even when the field theory is conformally invariant.
It can however be improved~\cite{Callan:1970ze} in the sense 
that terms $\Delta_{ab}$ that satisfy $\partial^a \Delta_{ab}=0$ 
and do not modify the conserved charges can be added in such a 
way that $T_{ab}$ is traceless when the theory is conformal.
See~\cite{Deser:1995ne} for a discussion in two-dimensional field theory.
It would be interesting to see whether the line of low-energy constants
$c_3=-2c_2$~\cite{Luscher:2004ib} plays a special role in this respect.

An observation of `practical' importance is that the linear combination 
\be
(T_{00}-T_{11})(t,y_1,\by)
 = 2\sigma \delta(\by-\bh(t,y_1))\left(1 + {\rm O}(\partial \bh)^4\right)
\la{eq:T00T11}
\ee
tracks the position of the string up to possibly quartic corrections.
This makes it the simplest operator to measure the mean 
square radius of the string fluctuations, which can be compared 
directly with the NLO formula obtain in~\cite{Gliozzi:2010zt}.

The rest of this paper is structured as follows. 
We start by studying the structure of the confining
string as seen by the energy-momentum tensor in section 
(\ref{sec:gff}). 
We then work out the constraints on three-point correlation functions
imposed by the open-closed string duality in section (\ref{sec:ocsd}).
The leading-order string formula~(\ref{eq:wlo}), generalized
to contain the contributions of excited states in the three-point function,
turns out to be consistent with the functional form in $r$ 
imposed by the closed-string spectral representation,
and we thereby identify the effective theory prediction for
the form factors of the closed strings.
In particular we find that the square radius of the ground
state closed string, defined in the standard way from the slope 
of its form factors at the origin, grows logarithmically with 
the length of the string $\beta$.
In section (\ref{sec:latEMT}) we give the explicit form 
of the energy-momentum tensor on the lattice in $d+1$ dimensions.
This allows us to interpret a recent high-accuracy calculation 
of the string width in numerical lattice gauge theory
in terms of matrix elements of the energy-momentum tensor.
In the rest of this paper, we work in Euclidean space,
and our sign conventions are as follows. In Minkowski space, 
the thermal expectation values of the diagonal components are 
$\<T_{00}\>=e$ and $\<T_{11}\>=p$ (respectively the energy density 
and pressure), while in Euclidean space $\<T_{00}\>=e$ 
and $\<T_{11}\>=-p$.

\section{Gravitational Form Factors of Closed Strings\la{sec:gff}}

In this section, we analyze how the transverse size of closed strings can be 
characterized.
In the pure SU($N$) gauge theory, the only conserved charges are energy and momentum.
Therefore, it is natural to measure the width of the string in terms of the 
distribution of these charges. 
While for the open string, the width can be probed directly in $x$-space,
it has to be defined initially in momentum space
through a form factor for the closed string:
the form factors with respect to the energy-momentum tensor $T_{\mu\nu}$
are the Fourier transforms of the energy and distributions. 
This simple relation between form factors and charge distribution applies 
because of the non-relativistic kinematics of the closed string, by which 
we mean that their transverse size is parametrically larger than their inverse mass.
By contrast, the electromagnetic form factors of the proton only 
correspond to the Fourier transform of charge and magnetization in the 
infinite-momentum frame~\cite{Burkardt:2000za}.

Here we will restrict ourselves to studying the form factors of states
that contribute to the Polyakov loop two-point function. These
states are translationally invariant in the longitudinal 
direction, therefore we restrict the momentum transfer to the transverse
directions. 
Furthermore, the closed string states are rotationally invariant, 
hence they have spin zero in $(d-1)$-dimensional space.

In order to exhaustively list the relevant form factors,
we decompose the full $(d+1)$-dimensional 
energy-momentum tensor into irreducible representations 
of $d$-dimensional space. %
The closed strings are stretched around a cycle of length $\beta$
in a spatial direction labeled $z$,
while the other spatial directions are labeled by $k,l=1,\dots ,(d-1)$.
Schematically, the decomposition takes the form
\be
\left(\begin{array}{c@{\quad}c}
T_{{0}{0}} & T_{{0}{k}}  \\
T_{{k}{0}} & T_{{k}{l}} \\
\hline
T_{{z}{0}} & T_{{z}{k}} 
\end{array}
\left|
\begin{array}{c}
T_{{0}{z}} \\
T_{{k}{z}} \\
\hline
T_{{z}{z}} 
\end{array}\right.
\right)\,.
\ee
In the following, we choose the normalization of states such that 
\be
\<\psi,P'|\psi,P\> = (2\pi)^{d-1}\delta^{d-1}(\boldsymbol{P}'-\boldsymbol{P}) 
         \cdot \beta\cdot  {2E_{\boldsymbol{P}}}.
\la{eq:norm}
\ee
The operator $T_{{z}{z}}$, which measures the stress in the ${z}$ direction,
is a scalar operator from the point of view of physics within an
${z}=\,$constant slice. Therefore, its matrix elements can be parametrized as
\be
\<\psi',P'|T_{{z}{z}}(0)  |\psi,P\> = 2 M\,M'\, f_3(\psi',\psi;\bq^2).
\ee
We use the standard notation $q=P'-P$, $\bar P =\half (P+P')$, and have 
accounted for the possibility that the mass of the final state $M'$
differs from the mass of the initial state $M$.

Secondly, we note that $(T_{{0}{z}},T_{{k}{z}})$ is a conserved vector 
from the point of view of a ${z}=\,$constant slice, 
if one restricts oneself to matrix elements between
states that are translationally invariant in the ${z}$ direction:
\be
\partial_{0} T_{{0}{z}} + \partial_{k} T_{{k}{z}} + \underbrace{\partial_{z} T_{{z}{z}}}_{=0} = 0.
\ee
We are thus dealing with the vector form factor of a scalar object,
hence (by analogy with the pion electromagnetic form factor),
\be
\<\psi',P'|T_{\mu{z}}(0)|\psi,P\> = M \bar{P}_\mu\, 
                f(\psi',\psi;\bq^2),\qquad \mu\neq z.
\ee
However, $T_{\mu{z}}$ is odd under the reflection $z\to-z$.
For matrix elements with  $P_z=P_{z}'=0$, this implies that $f$
must vanish identically~\footnote{Even in theories where the string is oriented,
as is the case in SU($N\geq 3$) theories, $T_{\mu z}$ cannot induce
transition between states created by $\re\tr{P}$ and $\im\tr{P}$
(which are respectively even and odd under the $z$-reflection symmetry), 
because such a matrix element vanishes by charge conjugation 
symmetry ($P=C$ for states coupling to the Polyakov loop).}.

Finally, the components of $T_{\mu\nu}$ not containing the index `$z$'
form a tensor with respect to the SO($d$) group. 
Taking again into account the fact that these components 
form a conserved tensor in the subspace of states invariant under 
translations along the $y$ direction, one finds that the general 
form of the matrix elements of $T_{\mu\nu}$ is 
\be
\,\<\psi',P'|T_{\mu\nu}(0)|\psi,P\> = -2\bar P_\mu \bar P_\nu f_1(\psi',\psi;\bq^2) 
        + 2(q_\mu q_\nu - q^2 \delta_{\mu\nu} )\,f_2(\psi',\psi;\bq^2),
\qquad \mu,\nu\neq z.
\la{eq:TabME}
\ee
Thus the transverse structure of the ground state of the string 
is characterized by a total of three form factors $\{f_i\}_{i=1}^3$.
The matrix elements 
\ba
\<\psi,P|T_{\mu\nu}(0)|\psi,P\>&=& -2 {P_\mu P_\nu},
\\
{\<\psi,P|T_{{z}{z}}(0)|\psi,P\>} &=& \beta \frac{\partial E^2_{\bp}}{\partial \beta},
\ea
determine the forward, diagonal matrix elements of $f_1$ and $f_3$,
\be
f_1(\psi,\psi;\boldsymbol{0})=1, \qquad 
f_3(\psi;\psi;\boldsymbol{0}) = 
\frac{\beta}{2M^2} \frac{\partial E^2_{\bp}(\beta)}{\partial \beta}.
\ee
The interpretation of these form factors is that 
$f_3$ measures the transverse distribution of longitudinal stress 
in the string, while $f_1$ measures the transverse distribution of 
energy. The form factor $f_2$ is somewhat less obvious to interpret.
For two states with momenta equal and opposite aligned along the direction $\hat1$
(Breit frame),
it describes the ability of $T_{22}$ to induce a transition between
these states per unit (momentum-transfer)$^2$ (this interpretation requires $d\geq3$). 
Indeed, in this kinematic configuration, $f_1$ does not contribute to the matrix 
element (\ref{eq:TabME}).

\section{Transverse Structure of open and closed strings\la{sec:ocsd}}

\begin{figure}
\centerline{\includegraphics[width=10.5 cm,angle=0]{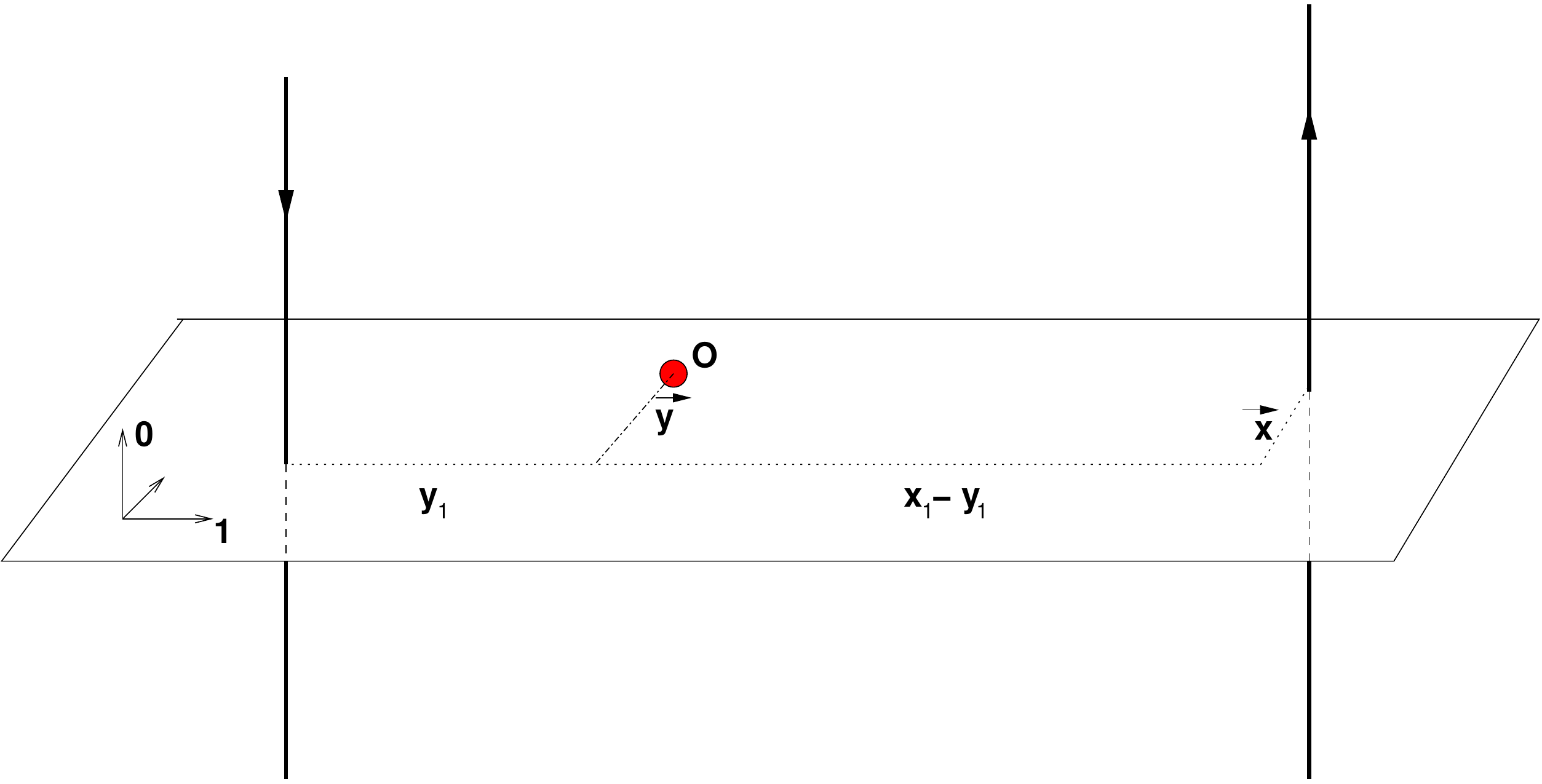}}
\caption{The geometry of the three-point function.}
\label{fig:3pt}
\end{figure}

The goal of this section is to derive the spectral representation 
of a three-point function where a local operator is used to probe
the structure of the confining string. We begin by recalling the 
spectral representation of the Polyakov loop two-point function.
The geometry of the Polyakov correlator is illustrated in \fig(\ref{fig:3pt}).

The open-string representation of the Polyakov loop two-point function
reads, setting $r^2\equiv x_1^2+\bx^2$ and with $w_n$ integer weights,
\be
\<P_0(x_1,\bx)\, P_0^*(0,\boldsymbol{0})\> 
= \sum_n w_n \, e^{-V_n(r) L}.
\ee
Upon introducing the matrix elements~\footnote{The matrix elements $b_n$ 
are related to the matrix elements $v_n$ defined in~\cite{Meyer:2006qx} by 
$b_n= v_n(\bp) \sqrt{{2\beta E^{(n)}_{\bp}}}$.}
\be
b_n\equiv 
\< {\rm vac}| P_0(0,\boldsymbol{0}) | n, \bp\> ,
\ee
the closed-string representation of the same correlation function reads
\ba
\<P_0(x_1,\bx)\, P_0^*(0,\boldsymbol{0})\>
&=& \frac{1}{\beta} \sum_n |b_n|^2  \int \frac{\ud^{d-1}\bp}{(2\pi)^{d-1}}\, 
 e^{i\bp\cdot\bx}\,\frac{e^{-E_n(\bp)x_1}}{2E_n(\bp)}
\\
&=& \sum_n |b_n|^2 \frac{r}{\beta M_n} \left(\frac{M_n}{2\pi r}\right)^{\frac{d}{2}}
K_{\frac{1}{2}(d-2)}(M_n r)
\\
&\sim & \sum_n \frac{|b_n|^2}{2\beta M_n}\, 
\left(\frac{M_n }{2\pi r}\right)^{\frac{d-1}{2}}
e^{-M_n r}\,.
\la{eq:2pt}
\ea
In the last line 
we have used the asymptotic form of the modified Bessel function,
$ K_\nu(x) \sim   e^{-x}\, \sqrt{\frac{\pi}{2x}}$;
the result is equivalent to using non-relativistic kinematics to begin with.
This expression dictates the functional dependence on $r$
of the Polyakov loop correlator.
As usual in deriving relations between open and closed strings,
the correlation function cannot be simultaneously dominated 
by a single open-string state \emph{and} a single closed-string 
state. Let $\beta$ be the length of the closed strings.
For $\beta\gg r$, a single open-string state dominates,
but O($\beta/r$) closed-string states contribute  in \eq(\ref{eq:2pt}).

Consider now the connected correlation function of 
a pair of Polyakov loops in the direction $\hat0$ and 
a local operator ${\cal O}$. Figure (\ref{fig:3pt}) illustrates 
the geometry of the correlator.
Its spectral interpretation in terms of open-string states
reads, for $\beta\gg \sigma^{-\frac{1}{2}}$, 
\be
\<P_0(x_1,\bx)\, {\cal O}(y_0,y_1,\by)\, P_0^*(0,\boldsymbol{0})\>
= \sum_n e^{-V_n(r)\beta}\, \<{\cal O}(y_1,\by)\>_n
\ee
In terms of closed-string states it can also be written as 
\be
\<P_0(x_1,\bx)\, {\cal O}(y_0,y_1,\by)\, P_0^*(0,\boldsymbol{0})\>
= \int \frac{{\ud}^{d-1}\bp'}{(2\pi)^{d-1}} e^{-i\bp'\cdot\bx}
\int \frac{{\ud}^{d-1}\bq}{(2\pi)^{d-1}} e^{-i\bq\cdot\by}
f(\bp',\bq, x_1, y_1),
\ee
where 
\be
f(\bp',\bq,x_1,y_1) = 
\int \ud^{d-1}\bx \, e^{i\bp'\cdot\bx}
\int \ud^{d-1}\by \, e^{i\bq\cdot\by} 
\<P_0(x_1,\bx)\, {\cal O}(y_0,y_1,\by)\, P_0^*(0,\boldsymbol{0})\>\,
\ee
is the correlation function in momentum space, which has a more 
natural interpretation from the closed-string point of view.
Here $\bp$ and $\bq$ have $d-1$ components.
Due to the translation invariance of the Polyakov loops along the 
$\hat0$ direction, $f$ has no dependence on $y_0$, which 
we therefore choose to be zero.

With the normalization of states given by \eq(\ref{eq:norm}),
we parametrize the matrix elements by
\ba
\<m,\bp'| {\cal O}| n,\bp\> &=& 2 M_m M_n \, F^{m,n}(\bar{\bp},\bq)\,,
\la{eq:FFME}
\\
\bp = \bp'- \bq,&\quad&  \bar{\bp}=\bp'-\half\bq = \half(\bp+\bp').
\ea
This parametrization is designed for dimension $(d+1)$ operators, 
for which $F^{m,n}$ is dimensionless.
We can now write the spectral representation of $f$,
\be
f(\bp',\bq,x_1,y_1) = 
\sum_{n,m} b_m \,\frac{e^{-E_m(\bp')(x_1-y_1)}}{2E_m(\bp')\beta}
2{M_m M_n}\,F^{m,n}(\bar{\bp},\bq) 
\frac{e^{-E_n(\bp)y_1}}{2E_n(\bp)\beta} \,  b_n^*,
~~(\bp = \bp'- \bq).
\la{eq:f}
\ee
Next we specialize to the case of a scalar operator
with respect to the symmetry group SO($d$) of a time-slice.
Examples thereof are $T_{00}$ or $T_{\mu\mu}$. 
We will return to the case of an operator with a more general
tensor structure in section (\ref{sec:T11}). 
Thus $F$ is a function of $\bq^2$ alone, hence 
\ba
\<P_0(x_1,\bx)\, {\cal O}(y_0,y_1,\by)\, P_0^*(0,\boldsymbol{0})\>
&=& 
\sum_{m,n} \frac{b_m b_n^*}{\beta^2}
 \int \frac{\ud^{d-1}\bq}{(2\pi)^{d-1}} e^{-i\bq\cdot\by}
\,2{M_m M_n}F^{m,n}(\bq^2) \, {\cal I}_{mn}(y_1,x_1-y_1,\bx,\bq),
\nn 
{\cal I}_{mn}(y_1,y_2,\bx,\bq) &=& 
\int \frac{\ud^{d-1} \bp'}{(2\pi)^{d-1}}
\frac{e^{-i\bp'\cdot\bx-E_m(\bp')y_2 - E_n(\bp'-\bq)y_1}}
{2E_m(\bp') 2E_n(\bp'-\bq)}.
\ea
We choose without loss of generality $\bx=0$. 
The quantity ${\cal I}$ is a massive one-loop integral,
\be
{\cal I}(y_1,y_2,0,\bq) = \int \frac{\ud\omega}{2\pi} e^{i\omega y_1}
\int \frac{\ud\omega'}{2\pi} e^{i\omega' y_2}
\int \frac{\ud^{d-1}\bp}{(2\pi)^{d-1}}
\frac{1}{\omega^2 +  (\bp-\bq)^2 + M_n^2}\,
\frac{1}{{\omega'}^2+\bp^2+M_m^2}.
\ee
This integral can be treated by standard techniques of 
quantum field theory, see for instance~\cite{Peskin:1995ev} p.327.
However we anticipate that 
non-relativistic kinematics is sufficient to study 
the long-distance behavior of the correlators 
(in the effective string theory, this will be guaranteed
as long as $\sigma \beta y_1\gg 1$),
\be
{\cal I}_{mn}(y_1,y_2,\boldsymbol{0},\bq) \sim
\left(\frac{M_m M_n}{y_2M_n + y_1M_m}\right)^{\frac{1}{2}(d-1)}\, 
\frac{\exp{-(M_m y_2+M_n y_1+\frac{\bq^2}{2}
\frac{y_1 y_2}{M_m y_1 + M_n y_2})}}{(2\pi)^{\frac{1}{2}(d-1)}\,2M_m\cdot 2M_n},
\ee
Therefore, with $y_2 \doteq x_1-y_1$, 
\ba
\<P_0(x_1,\boldsymbol{0})\, {\cal O}(y_0,y_1,\by)\, P_0^*(0,\boldsymbol{0})\>
&\sim&
\sum_{m,n} \frac{b_m b_n^* }{\beta^2}
\left[\frac{M_m M_n}{y_2M_n + y_1M_m}\right]^{\frac{d-1}{2}}
\frac{e^{-(y_2M_m+y_1M_n)}}{2(2\pi)^{\frac{1}{2}(d-1)}}
\cdot \la{eq:3pt}\\
&& ~~\cdot
\int \frac{\ud^{d-1}\bq}{(2\pi)^{d-1}} e^{-i\bq\cdot\by} F^{m,n}(\bq^2) 
e^{-\frac{\bq^2}{2} \frac{y_1 y_2}{M_my_1+M_n y_2}}.
\nonumber
\ea
This expression dictates the leading-order 
functional dependence on $x_1$ and $y_1$ of the three-point function
that the effective string theory must respect.

Expression (\ref{eq:3pt}) can be viewed as a distribution in $\by$.
The quantity we will confront with a prediction from the 
effective string theory is its second moment at $x_1\doteq r$,
\be
w^2(r,\beta,y_1) \equiv \frac{ \int \ud^{d-1}\by \, \by^2\,
\<P_0(r,\boldsymbol{0})\, {\cal O}(y_0,y_1,\by)\, P_0^*(0,\boldsymbol{0})\>}
{\int \ud^{d-1}\by \, 
\<P_0(r,\boldsymbol{0})\, {\cal O}(y_0,y_1,\by)\, P_0^*(0,\boldsymbol{0})\>}.
\la{eq:defw2}
\ee
Based on (\ref{eq:3pt}), we obtain
\be
w^2(r,\beta,y_1) = -2(d-1)\frac{d}{d\bq^2}
\log\Big\{\sum_{m,n}b_m b_n^*
{\txts\left[\frac{M_m M_n}{y_2M_n+y_1M_m}\right]^{\frac{d-1}{2}}}\!\!
e^{-(y_2 M_m+y_1 M_n+\frac{\bq^2}{2}\frac{y_1y_2}{M_my_1+M_ny_2})} F^{m,n}(\bq^2)\Big\}_{\bq=0}.
\la{eq:2ndA}
\ee
where $y_2\doteq r-y_1$. At $y_1=y_2=\frac{r}{2}$, the expression simplifies slightly,
\be
w^2(r,\beta,{\txts\frac{r}{2}}) = -2(d-1)\frac{d}{d\bq^2}
\log\Big\{\sum_{m,n}b_m b_n^*  
\mu_{mn}^{\frac{1}{2}(d-1)}
e^{-\frac{r}{2}(M_m+M_n+\frac{1}{2}\frac{\bq^2}{M_m+M_n})} F^{m,n}(\bq^2)\Big\}_{\bq=0},
\la{eq:2ndB}
\ee
where $\mu_{mn}$ is the reduced mass of $M_m$ and $M_n$ defined by
 $\mu_{mn}^{-1}=M_m^{-1}+M_n^{-1}$.

\subsection{Effective string theory prediction}
On the other hand, Allais and Caselle~\cite{Allais:2008bk} 
(see also the recent two-loop result~\cite{Gliozzi:2010zt}) obtained within the 
effective bosonic string theory the leading-order result,
for $x_1 = 2y_1 = r$,
\be
w^2_{lo}(r,\beta,{\txts\frac{r}{2}}) = 
\frac{d-1}{2\pi\sigma} \log\frac{r}{r_0} + \frac{1}{\pi\sigma}
\log \frac{Z_0^2(\beta,r)}{Z_0(2\beta,r)}.
\la{eq:ac}
\ee
Written in this form, it is clear that the second term 
can be interpreted as a difference of free energies.

\subsection{Transverse structure of the ground state of the closed string}
The limit  $y_1\gg \beta$  is most transparent from the 
closed string point of view, since the correlation function is then 
dominated by the closed-string ground state. 
Eq. (\ref{eq:2ndA}) yields in that limit
\be
w^2(r,\beta,y_1) 
= -2(d-1)\left[-\frac{y_1(r-y_1)}{2\sigma\beta r}+ \frac{(F^{0,0})'}{F^{0,0}} \right]
\stackrel{y_1=\frac{r}{2}}{=} 
 -2(d-1)\left[-\frac{r}{8\sigma\beta}+ \frac{(F^{0,0})'}{F^{0,0}} \right],
\la{eq:gen}
\ee
where we have used the leading order relation $M_n=\sigma L$.
The form factors are now evaluated at $\bq^2=0$, and the prime denotes
differentiation with respect to $\bq^2$.
In the regime $y_1\gg \beta$, the effective string expression (\ref{eq:ac}) 
behaves as 
\be
w^2_{lo}(r,\beta,{\txts\frac{r}{2}}) = 
\frac{d-1}{2\pi\sigma}\log\frac{\beta}{4r_0} 
+\frac{d-1}{4\beta\sigma}r + {\rm O}(e^{-2\pi r/\beta}).
\la{eq:ac2}
\ee
It is consistent with the general expression (\ref{eq:gen}) 
derived from the spectral representation of the correlator.
The linear term turns out to agree automatically between the two
expressions. From the closed-string point of view,
this term is essentially a kinematic effect; we will return to its
significance in the open-string interpretation of the three-point function.

The rms radius of the closed string, defined in the standard way from the 
derivative of the form factor at the origin, can be identified 
with the $r$-independent term,
\be
\<r^2\>_{\rm closed} \equiv -\frac{2(d-1)}{F^{0,0}(\boldsymbol{0})}
\left.\frac{dF^{0,0}(\bq^2)}{d\bq^2}\right|_{\bq=0}
= \frac{d-1}{2\pi\sigma}\log\frac{\beta}{4r_0}.
\ee
This term thus measures the logarithmic 
broadening of the \emph{closed} string with its length $\beta$.
The prefactor is the same as for the open string, but the UV length scale
appearing inside the logarithm is four times larger than in the open-string case.

When $r\gg\beta$, the open-string ensemble is at finite temperature $1/\beta$.
The local operator then probes the profile of the open-string states,
averaged over with the Boltzmann weight. Equation (\ref{eq:gen}) 
shows that the profile at mid-string grows linearly with the length 
of the open string~\cite{Allais:2008bk}. 
This linear rise is likely due to the fact that O($r/\beta$)
open-string states contribute to the correlation function
when $r\gg\beta$, and the width results from a stochastic
superposition of these contributions.
A linear increase is in fact nothing exotic,
since for a screened potential $V(r)\sim e^{-mr}$, 
the profile goes like $e^{-m\sqrt{(r/2)^2+\by^2}}$,  and hence
the mean square radius is given by $(d-1)\frac{r}{2m}$ for large $r$.

\subsection{Interpretation of excited closed-string contributions}

Both the general expression (\ref{eq:2ndB}) and the bosonic string 
formula (\ref{eq:ac}) can be expanded in a series of exponentials
that fall off increasingly fast. We require that the 
coefficients of these exponentials match.

In the following, we use the leading-order relation between the matrix elements
$b_n$ and the multiplicity (integer) factors $w_n$, 
\be
\left|\frac{b_n}{b_0}\right|^2 \stackrel{l.o.}{= } \frac{w_n}{w_0},
\ee
we choose them to be real
and use the fact that $w_0=1$. We recall the values $w_1=d-1$ and 
$w_2 = 1+(d-1)+\half (d+1)(d-2)$~\cite{Luscher:2004ib},
and also define $\Delta M_n \equiv M_n - M_0$.
We start by analyzing the leading 
correction to (\ref{eq:ac2}), which comes solely from $Z_0(2\beta)$,
\be
w^2_{lo}(r,\beta,{\txts\frac{r}{2}})
 \supset -\frac{w_1}{\pi\sigma}\, e^{-\frac{\Delta M_1 r}{2}}.
\ee
Expanding (\ref{eq:3pt}), one finds that the O($r$) term cancels out
automatically.  From the O($r^0$) term, we obtain the consistency condition
\be
{2(d-1)}\, 
\frac{d}{d\bq^2}\left[\frac{\re F^{1,0}(\bq^2)}{F^{0,0}(\bq^2)}\right]_{\bq=0}
 = \frac{\sqrt{w_1}}{2\pi\sigma},
\la{eq:F10}
\ee
which dictates the strength of the off-diagonal matrix element 
between the lightest two string states at small momentum transfer.

We now turn to the term of order $e^{-\Delta M_1r}$, which is 
of precisely the same order as $e^{-\frac{1}{2}\Delta M_2 r}$ for the leading order
spectrum. This time, both $Z_0(2\beta)$ and $Z_0^2(\beta)$ contribute and we find 
\be
w^2_{lo}(r,\beta,{\txts\frac{r}{2}})
 \supset \frac{2w_1-w_2+ \half w_1^2}{\pi\sigma}\, e^{-\Delta M_1 r},
\la{eq:DM1lo}
\ee
while from the general expression, we extract
\be
w^2(\beta,r,{\txts\frac{r}{2}})\supset -2(d-1)e^{-\Delta M_1 r}\frac{d}{d\bq^2}
\left[w_1 \frac{F^{1,1}}{F^{0,0}}
+ 2\sqrt{w_2} \frac{\re F^{2,0}}{F^{0,0}} 
- 2 w_1 \left(\frac{\re F^{1,0}}{F^{0,0}}\right)^2\right]_{\bq=0}.
\la{eq:DM1}
\ee
The comparison of \eq(\ref{eq:DM1lo}) and (\ref{eq:DM1}) yields predictions for 
the form factor at small momentum transfer.
By generalizing $w^2$ to values of $y_1\neq y_2$, one could disentangle
$F^{2,0}$ from $F^{1,1}$ and obtain separate predictions for these form factors.
In this way, a sequence of predictions are obtained for the form factors between 
low-lying states.

\subsection{Three-point function with a non-scalar probe operator\la{sec:T11}}

We now come back to (\ref{eq:f}) in the case of an operator with a more complicated 
tensor structure. Consider the case of  ${\cal O}= T_{11}$.
Recall that direction `1' plays the role of time from the point 
of view of the closed strings.
Then we replace \eq(\ref{eq:FFME}) by (see \eq(\ref{eq:TabME}))
\be
\<m,\bp'| {\cal O}| n,\bp\>  =
{\txts\frac{1}{2}} [E_m(\bp')+E_n(\bp)]^2 f^{m,n}_1(\bq^2) - 2\bq^2 f^{m,n}_2(\bq^2)
\ee
In this case we can write 
\ba
\<P_0(x_1,\bx)\, {\cal O}(y_0,y_1,\by)\, P_0^*(0,\boldsymbol{0})\> &=&
\sum_{m,n} \frac{b_m b_n^*}{\beta^2}
 \int \frac{\ud^{d-1}\bq}{(2\pi)^{d-1}} e^{-i\bq\cdot\by}\cdot 
\\
&&
\left[\half f_1^{m,n}(\bq^2) {\txts(\frac{\partial}{\partial y_1}+\frac{\partial}{\partial y_2})^2}
-2\bq^2 f_2^{m,n}(\bq^2) \right]{\cal I}_{mn}(y_1,y_2,\bx,\bq).
\nonumber
\ea
where $y_2$ is set to $x_1-y_1$ at the end.
We now note that at leading order for large $r=2y_1=2y_2$, 
\be
{\txts(\frac{\partial}{\partial y_1}+\frac{\partial}{\partial y_2})^2}
{\cal I}_{mn}(y_1,y_2,\boldsymbol{0},\bq) \sim (M_m+M_n)^2 
{\cal I}_{mn}({\txts\frac{r}{2}},{\txts\frac{r}{2}},\boldsymbol{0},\bq).
\ee
Hence to leading order the preceding analysis still applies, with
the substitution
\be
F^{m,n}(\bq^2) \to f_1^{m,n}(\bq^2) -\frac{\bq^2}{M_mM_n}f_2^{m,n}(\bq^2).
\ee
A special feature of the operator $T_{11}$ is that 
\be
f^{m,n}_1(\boldsymbol{0}) = \delta_{mn},
\ee
since the states $|n,\bp\>$ are energy eigenstates. 
In particular, \eq(\ref{eq:F10}) simplifies slightly to 
\be
-2(d-1)\re (f_1^{1,0})'(\boldsymbol{0}) = \frac{\sqrt{w_1}}{2\pi\sigma}.
\ee
It is interesting that, with our normalization of states (\ref{eq:norm}),
the transition form factor is independent of $\beta$.

\section{Lattice definition of the energy-momentum tensor 
          in $(d+1)$-dimensions\la{sec:latEMT}}
In this section, we derive the lattice form of the energy-momentum 
tensor in $(d+1)$-dimensional SU($N$) gauge theory.
Our main motivation is that these operators have been mostly 
studied in the $d=3$ case, but recently there has been extensive 
work on strings in $d=2$ SU($N$) 
gauge theories~\cite{Athenodorou:2008cj,Gliozzi:2010zv}. 
This preparatory work will help us interpret those results.

We will follow the treatment~\cite{Meyer:2007fc}
and generalize it to $d$ dimensions.
The idea is to identify the operators whose expectation
value yield the thermodynamic energy density and pressure.
We start from the Wilson action~\cite{Wilson:1974sk} 
on an anisotropic lattice~\cite{Engels:1980ty},
\be
S_{\rm g}= \sum_x \betas \Ss(x) + \betat \St(x).
\ee
The action has two bare parameters, $\betas$ and $\betat$, 
and there are two `renormalized' parameters, the 
spatial lattice spacing $\as$ and the 
renormalized anisotropy $\xi=\as/\at$.
At the isotropic point, $\betas=\betat=\beta$ (not to be 
confused with the symbol used for the closed-string length in the previous
sections).
The function $\Ss$ and $\St$ of the link variables $U_\mu(x)$ 
contain exclusively spatial and temporal Wilson loops, respectively.
The partition function $Z$ depends on $\betas$, $\betat$ 
and the lattice dimensions, $\Nt\cdot\Ns^d$.
The latter are related to its physical size by $L=\Ns \as$, 
$L_0= 1/T = \Nt \at$.
We define the renormalized quantity $\overline Z$ by
\be
\log \overline{Z}(\betas,\betat,\Ns,\Nt)
=\log Z(\betas,\betat,\Ns,\Nt) 
- \frac{\Nt}{\Nt^{\rm ref}} \log Z(\betas,\betat,\Ns,\Nt^{\rm ref}).
\ee
The conditions that $\overline{Z}$ does not depend 
on $\as$ or on the anisotropy $\xi$ translate respectively into
\ba
\frac{L\partial\log\overline{Z}}{\partial L} + 
\frac{L_0\partial\log\overline{Z}}{\partial L_0} 
&=& 
-\sum_{x} \frac{\partial\betas}{\partial\log\as}\<\Ss\> 
+ \frac{\partial\betat}{\partial\log\as} \<\St\>\,,
\\
\frac{L_0\partial\log\overline{Z}}{\partial L_0} 
&=&
\sum_{x} \frac{\betas}{\partial\log\xi}\<\Ss\>
+ \frac{\partial\betat}{\partial\log\xi}\<\St\>,
\ea
where it is understood that the expectation values of $\Ss$
and $\St$ on the $\Nt^{\rm ref}\cdot \Ns^d$ lattice are subtracted.
We then recall the thermodynamic
definitions of energy density and pressure,
\be
e =-\frac{1}{L_0L^d}\,
\frac{L_0\partial\log\overline{Z}}{\partial L_{0}}\,,
\qquad
p=\frac{1}{dL_0L^d}\, \frac{L\partial\log\overline{Z}}{\partial L}.
\ee
With these definitions, we obtain at the isotropic point $\xi=1$,
\ba
a^{d+1}(e - dp) &=&  \frac{\partial\betas}{\partial\log\as}\<\Ss\>
                   +\frac{\partial\betat}{\partial\log\as}\<\St\>,
\la{eq:e-dp}
\\
\frac{d}{d+1}a^{d+1}(e+p) &=& 
-\left(\frac{\partial\betas}{\partial\log\xi}
+\frac{1}{d+1}\frac{\partial\betas}{\partial\log\as}\right) \<\Ss\>
-\left( \frac{\partial\betat}{\partial\log\xi} 
+ \frac{1}{d+1}\frac{\partial\betat}{\partial\log\as} \right) \<\St\>\,.
\la{eq:e+p}
\ea

On the other hand, from the definition of the stress-energy tensor,
we expect that
\be
\<\theta\>\equiv \<T_{\mu\mu}\> = e-dp,\qquad  \<T_{00}\> =e .
\ee
We also define 
\be
\theta_{\mu\nu} = T_{\mu\nu} - \frac{1}{d+1}\delta_{\mu\nu}\theta,
\ee
so that in particular
\be
\<\theta_{00}\> = \frac{d}{d+1}(e+p).
\ee
Since \eq(\ref{eq:e-dp}) and (\ref{eq:e+p}) hold at every temperature, 
we infer that 
\ba
a^{d+1}\theta &=& \frac{\partial\betas}{\partial\log\as}\Ss
                 + \frac{\partial\betat}{\partial\log\as}\St,
\la{eq:th}
\\
a^{d+1} \theta_{00} &=& 
-\left(\frac{\partial\betas}{\partial\log\xi}
+\frac{1}{d+1}\frac{\partial\betas}{\partial\log\as}\right) \Ss
-\left( \frac{\partial\betat}{\partial\log\xi} 
+ \frac{1}{d+1}\frac{\partial\betat}{\partial\log\as} \right) \St\,.
\la{eq:th00}
\ea

Recall that the magnetic field has $\frac{d(d-1)}{2}$ components,
while the electric field has $d$ components. The lattice action 
can be expressed in terms of these fields,
\be
\Ss=\frac{\as^{4}}{N_c}\tr\{\boldsymbol{B}^2\}\,,
\qquad
\St=\frac{\as^2\at^2}{N_c}\tr\{\boldsymbol{E}^2\}\,.
\ee
An important observation is now that at the isotropic point $\xi=1$, 
the operators
\be
\St - {\txts\frac{2}{d-1}}\Ss
\qquad 
{\rm and}
\qquad
\St + \Ss
\ee
belong to irreducible representations of the cubic group in $(d+1)$ 
dimensions~\footnote{For instance, in $d=2$ the operators 
$E_x^2-B^2$ and $E_y^2-B^2$ span an irreducible representation of the cubic
group, and $\St-2\Ss$ is proportional to the sum of the two.}. 
Since in both cases there is no other gauge-invariant operator of dimension 
$(d+1)$ in the same representation, both of them renormalize multiplicatively.

\subsection{The case $d=2$}

Since the $d=3$ case is well known~\cite{Engels:1980ty,Meyer:2007fc}, 
we focus here on the $d=2$ case.
The $d=2$ theory is super-renormalizable, which leads to considerable
simplifications. At treelevel on the anisotropic lattice, we have the 
following expressions for the bare parameters in terms of the renormalized 
ones,
\be
\betas = \frac{2N_c}{g^2\as}\frac{1}{\xi},\qquad
\betat=\frac{2N_c}{g^2\as}\xi
\qquad {\rm (treelevel)}.
\ee
Hence
\ba
\frac{\partial\betas}{\partial\log\as} \simeq -\betas,&\qquad&
\frac{\partial\betat}{\partial\log\as} \simeq -\betat,
\\
\frac{\partial\betas}{\partial\log\xi}\simeq -\betas,&\qquad&
\frac{\partial\betat}{\partial\log\xi}\simeq \betat.
\ea
Inserting these expressions into \eq(\ref{eq:th00}),
we get the following treelevel expressions at the isotropic point,
\ba
a^3 \theta &=& 
-\beta (\Ss+\St)\,,
\\
a^3\theta_{00} &=& \frac{2}{3}\beta (2\Ss - \St)\,.
\ea
Since we already know that these linear combinations 
renormalize multiplicatively 
(see the remarks at the end of the last section), 
the full expressions for $\theta$ and  $\theta_{00}$ read
\ba
\theta &=& 
\dBdloga (\Ss+\St)\,,
\la{eq:th_d=3}
\\
\theta_{00} &=& \frac{2}{3}\beta  Z(\beta) (2\Ss - \St)\,,
\la{eq:th00_d=3}
\ea
with $Z$ of the form $Z(\beta)=1 + {\rm O}(\beta^{-1})$ and 
$\dBdloga = -\beta (1+{\rm O}(\beta^{-1}))$. 
Now comparing these expressions with \eq(\ref{eq:th}) and (\ref{eq:th00}),
we obtain at $\xi=1$ the relations
\ba
-\frac{\partial(\betas+2\betat)}{\partial\log\xi} &=& \dBdloga,
\\
\frac{\partial(\betat-\betas)}{\partial\log\xi} &=& 2\beta Z(\beta) .
\ea
Combining (\ref{eq:th_d=3}) and (\ref{eq:th00_d=3}),
\be
a^3T_{00} = \Ss(\frac{4}{3}\beta Z(\beta)+\frac{1}{3}\frac{\partial\beta}{\partial\log a})
+\St(-\frac{2}{3}\beta Z(\beta)+ \frac{1}{3}\frac{\partial\beta}{\partial\log a}).
\ee
By Euclidean symmetry, one then obtains also the expression for the diagonal stress operator,
\be
T_{xx} = S_{0y}(\frac{4}{3}\beta Z(\beta)+\frac{1}{3}\frac{\partial\beta}{\partial\log a})
+ (S_{0x}+S_{xy}) (-\frac{2}{3}\beta Z(\beta)+ \frac{1}{3}\frac{\partial\beta}{\partial\log a}),
\ee
and similarly for $T_{yy}$.

In summary, we have derived the lattice expressions for the renormalized diagonal components
of the energy-momentum tensor. A simplification of the $d=2$ case over the usual $d=3$ case
is that the one-loop quantum corrections to $Z$ and $\dBdloga$ amount to O($a$) effects, 
and the two-loop effects would amount to O($a^2$) corrections. The latter 
are parametrically of the same order as the usual O($a^2$) cutoff effects 
that are expected
to occur in lattice gauge theory. For that reason, a one-loop computation
is sufficient to yield a fully renormalized energy-momentum tensor.

\subsubsection{Application: width of the confining string}

In~\cite{Gliozzi:2010zv}, 
the width of the string, stretched between two static charges 
separated by a distance $r$ along the $x$ direction,
was extracted from the measurement of the $P_{0x}=-S_{0x}+{\rm cst}$ 
plaquette expectation value at the midpoint $x=r/2$ 
(we now specialize to the case of the Wilson action;
the additive constant drops out when subtracting the vev of the plaquette).
We now interpret this result in terms of the energy-momentum tensor
derived above.

Working at treelevel, 
\be
\left(\begin{array}{c} T_{00} \\ T_{xx} \\ T_{yy} \end{array}\right)
= \frac{\beta}{a^3}
\left(\begin{array}{c@{\quad}c@{\quad}c}    
+1 & -1 & -1 \\
-1 & -1 & +1 \\
-1 & +1 & -1 
  \end{array}\right)
\left(\begin{array}{c} S_{xy}\\S_{0x}\\S_{0y}\end{array}\right)\,.
\la{eq:mat}
\ee
Inverting the matrix, one finds that 
\ba
S_{xy} &=& -\frac{a^3}{2\beta} (T_{xx}+T_{yy}),  
\\
S_{0x} &=& -\frac{a^3}{2\beta} (T_{00}+T_{xx}),
\\
S_{0y} &=& -\frac{a^3}{2\beta} (T_{00}+T_{yy}).
\ea
Now we use the general sum rules
\ba
\frac{\<\psi|\int d^d\boldsymbol{x} T_{00}(x)|\psi\>}{\<\psi|\psi\>} &=& E,
\\
\frac{\<\psi|\int d^d\boldsymbol{x} T_{xx}(x)|\psi\>}{\<\psi|\psi\>} &=& 
L_x\frac{\partial E}{\partial L_x}.
\ea
Here $L_x$ represents an external parameter that $E$ depends on.
Thus, for a string of length $r$ along the $x$-direction,
\ba
-\frac{\<\psi|\beta\sum_{\boldsymbol x} S_{xy}(x)|\psi\>}{\<\psi|\psi\>} &=& 
\half ar\frac{\partial E}{\partial r}, 
\\
-\frac{\<\psi|\beta\sum_{\boldsymbol x} S_{0x}(x)|\psi\>}{\<\psi|\psi\>} &=& 
\half a (E+r\frac{\partial E}{\partial r}),
\la{eq:S0x}
\\
-\frac{\<\psi|\beta\sum_{\boldsymbol x} S_{0y}(x)|\psi\>}{\<\psi|\psi\>} &=& 
\half a E.
\ea
These can be viewed as the $d=2$ version of the 
Michael sum rules~\cite{Michael:1995pv}.
For a long string, where $E\propto r$, we expect the various plaquettes
(summed over a time-slice) to come in the fractions
\be
\<S_{xy}\>: \<S_{0x}\>:\<S_{0y}\> = \half: 1 :\half.
\ee

We finish with a numerical application of \eq(\ref{eq:S0x})
based on the data of~\cite{Gliozzi:2010zv}.
For the rest of this section, we set the lattice spacing to unity.
The profile obtained in~\cite{Gliozzi:2010zv}
from the $S_{0x}$ operator is to a good approximation Gaussian, with 
\be
\int_{-\infty}^{\infty}\ud y\, A\exp(-\half y^2/R^2) = \sqrt{2\pi}\cdot A\cdot R.
\ee

From Fig. (2) of~\cite{Gliozzi:2010zv}, one reads off $R\approx\sqrt{12.1}$, 
and from Fig. (1), $A\approx 0.00038$. 
Thus the left hand side of \eq(\ref{eq:S0x}) roughly amounts to 
\be
\sqrt{2\pi}\beta A R r. 
\ee 
If we neglect the quark self-energies and the string corrections, 
$E\approx \sigma r$ and the RHS amounts to 
\be
\sigma r.
\ee
Numerically~\cite{Gliozzi:2010zv}, after simplifying the common factor $r$, 
we have $LHS\approx0.030$ and $RHS\approx 0.026$. Given the approximations we have made, 
in particular the neglect of the quark self-energies and the use of the treelevel
renormalization factors for the plaquette, the agreement is satisfactory.

Based on the remarks of section (\ref{sec:coupling}), we expect all three plaquettes
to yield the same string profile, to leading order in the string fluctuations:
each of them contains a piece of either the energy density $T_{00}$
or the longitudinal stress $T_{xx}$. This is indeed what the authors 
of~\cite{Gliozzi:2010zv} observed. We would however  expect $T_{yy}$, 
whose expressions in terms of plaquettes can be read off \eq(\ref{eq:mat}), 
to yield a different profile.

\subsection{The case $d=3$}

It was numerically observed a long time ago~\cite{Sommer:1988aa} 
in $d=3$ dimensions and for the gauge group SU(2) that 
the trace anomaly operator $T_{\mu\mu}$ yields a large 
string width that grows with $r$ in the observable (\ref{eq:defw2}). 
The linear combination
$3T_{00}-T_{11}-(T_{22}+T_{33})$ was found to yield a smaller value, 
and no clear evidence for a growing width could be seen. 
According to our conjecture for the coupling of the stress-energy
tensor to the string, this operator should measure the same
width at leading order for a long enough string, but in 
the range $r\sqrt{\sigma}<2$ reached in the study 
the corrections could be significant.

A few years later, a new numerical study was carried out
\cite{Green:1996be,Pennanen:1997qm} in the same theory.
The authors considered the operators 
$T_{\mu\mu}$ and $\left(T_{00}+T_{11}-(T_{22}+T_{33})\right)$,
which yield similar profiles, as expected from the leading terms in \eq(\ref{eq:T00exp})
and (\ref{eq:T11}).
The operator that the authors call the `transverse energy' is proportional to $T_{00}-T_{11}$, 
and there is some evidence that the profile measured with this linear combination
is indeed different, as we would expect based on the arguments of 
section (\ref{sec:coupling}). The reader is reminded that 
we are using Euclidean conventions here,
see the comment at the end of section (\ref{sec:coupling}).

\section{Conclusion}
In this paper we have analyzed the transverse structure of the 
confining string in non-Abelian gauge theories.
By postulating a particular coupling of the stress-energy tensor
to the string degrees of freedom, we obtained a prescription
for how to compare the string profile obtained from local
gauge theory operators with worldsheet observables.
We then derived the closed string representation of the 
three-point function from which the string profile can be extracted.
For this purpose we first enumerated the gravitational form factors
that characterize the string profile. The functional form of the 
leading-order prediction for the string's square width is then found 
to be in agreement with the closed-string spectral representation.
Most importantly, we showed that the square radius of the ground-state
closed string, defined from the slope of its form factor, grows 
logarithmically with the length of the string, just as the square
radius of the open string does. We also obtained a prediction for
the transition form factor between the ground and the 
first excited state.

A number of open questions remain.
It is not quite clear yet what role the ambiguity in the form of the 
worldsheet energy-momentum tensor (canonical vs. improved) plays beyond 
the quadratic order.
It is not clear either to the author how exactly a generic 
local gauge-invariant operator couples to the worldsheet fields.
Finally, we wish to comment on the prospects of fully characterizing
the QCD string's structure. In the analysis of hadron structure, 
Generalized Parton Distributions have provided a powerful 
way to characterize the structure of a relativistic bound state
such as the proton (see~\cite{Belitsky:2005qn} for a review of the subject).
Their moments in the longitudinal momentum fraction 
are given by the form factors of twist-two operators
and are thus computable in the Euclidean theory~\cite{Hagler:2009ni}.
These moments can be interpreted
as Fourier transforms of the transverse distribution 
of partons~\cite{Burkardt:2000za}.
The higher the dimension of the operator, the higher the longitudinal
momentum fraction of the partons that it is measuring the 
transverse distribution of.

It is a fascinating program to think about an analogous 
comprehensive way of characterizing the structure of the confining string.
Here there is no need to go to the infinite-momentum frame, which 
leads to kinematic simplifications in the proton case, because the string
is parametrically heavy compared to its transverse width. This warrants
the interpretation of form factors as the Fourier transforms of `parton'
densities. By analogy with the analysis of proton structure~\cite{Ji:1996nm},
a rational for which tower of operators to concentrate on 
may be provided by a
`Deeply Virtual Graviton Scattering' gedankenexperiment.
Higher-dimensional operators will presumably 
correspond to probing the transverse distribution of `gravipartons'
which carry a higher fraction of the string's energy. By the well-known
arguments, we would expect to find smaller transverse radii 
for these operators.

\acknowledgments{
This paper is based on an invited contribution to the workshop
``Confining Strings'' in Trento, 5-9 July 2010. I thank Ofer Aharony, 
Barak Bringoltz and Michael Teper for organizing a very stimulating and 
fruitful workshop, which was supported 
by the European Community - Research Infrastructure Action under the
FP7 ``Capacities'' Specific Programme, project ``HadronPhysics2''.
}

\bibliography{/home/meyerh/CTPHOPPER/ctphopper-home/BIBLIO/viscobib.bib}

\end{document}